\documentclass[twocolumn,floatfix,showpacs,preprintnumbers,amsmath,amssymb]{revtex4}
\usepackage{graphicx}
\usepackage{dcolumn}
\usepackage{bm}
\usepackage{color}
\usepackage{tabularx}
\usepackage{rotating}

\begin{document}
\title{ Structure and magnetic properties of nanocrystalline PrCo$_3$}

\author {K. Younsi}
\affiliation{CMTR, ICMPE, UMR7182, CNRS $-$ Universit\'e Paris 12, \\ 2-8 rue Henri Dunant F-94320 Thiais,  France}

\author {V. Russier}
\affiliation{CMTR, ICMPE, UMR7182, CNRS $-$ Universit\'e Paris 12, \\ 2-8 rue Henri Dunant F-94320 Thiais,  France}

\author {L. Bessais\footnote[1]{Author to whom correspondence should be addressed; electronic mail:
bessais@glvt-cnrs.fr}}
\affiliation{CMTR, ICMPE, UMR7182, CNRS $-$ Universit\'e Paris 12, \\ 2-8 rue Henri Dunant F-94320 Thiais,  France}

\date{\today}
\begin{abstract}
The structure and magnetic properties of nanocrystalline PrCo$_3$ prepared by high-energy milling technique have been investigated by means of X-ray diffraction using the Rietveld method coupled to Curie temperature and magnetic measurements. 
The as-milled samples were subsequently annealed in temperature range from 750 to 1050$^\circ$C for 30\,min to optimize the extrinsic properties. 
From x-ray studies of magnetic aligned samples, the magnetic anisotropy of this compounds is found uniaxial. The Curie temperature is 349\,K and no saturation reached at room temperature for applied field of 90\,kOe. The coercive field of 55\,kOe and 12\,kOe measured at 10 and 293\,K respectively is obtained after annealing at 750$^\circ$C for 30\,min suggests that nanocrystalline PrCo$_3$ are interesting candidates in the field of permanent magnets. 
We have completed this experimental study by simulations in the micromagnetic framework in order to get a qualitative picture of the microstructure effect on the macroscopic magnetization curve. From this simple model calculation, we can suggest that the after annealing the system behaves as magnetically hard crystallites embedded in a weakly magnetized amorphous matrix. 

\end{abstract}

\maketitle
\section{Introduction}
The intermetallic compounds, RCo$_3$ (R= rare earth element) have attracted great interest caused mainly by their wide applications as permanent magnets. These compounds present excellent magnetic properties such as large magnetocrystalline anisotropy and important saturation magnetization at room temperature. This interesting  magnetic performances of the intermetallic compounds formed by alloying rare earth and $3d$ transition metals are due to the combination of the complementary characteristics of $3d$-itinerant and $4f$-localized magnetism.
The coercivity is one of the most important extrinsic properties of a permanent magnet. This properties 
+depends on the microstructure of the material. 
Recently nanocrystalline compounds have been extensively investigated, their aim is to find the appropriate nanocrystalline state for a highest coercive field.

The most employed process to prepare these materials are rapid quenching by melt spinning \cite{gdy04} and mechanical alloying \cite{mdf96}. More recently high-energy milling technique, 
has also been used to produce high coercivity nanocrystalline powders. Rivera-G\'omez \textit{et al} are reported in \cite{rem09} a high coercivity for PrCo$_5$ compounds milling and annealed at 800$^\circ$C for 1\,min with $H_\text{c}= 17.3$\,kOe. A high coercivity of Sm$_2$Co$_{17}$ nanocrystalline compounds has been obtained by using this method \cite{cmhh00} where $H_\text{c}= 9.6$\,kOe after milling and annealed at 800$^\circ$C for 30\,min. 

In this paper, we report the elaboration of  nanocrystalline PrCo$_3$ by high-energy milling, and their structural and magnetic properties in amorphous state and after recrystallization. Hard magnetic properties of the as milled PrCo$_3$ have been improved by a controlled nanocrystallization, which optimize the size of grains in order to obtain a high coercivity. 

We also propose a simple model for the system after annealing based on a simple lattice of crystallites
embedded in an amorphous matrix. The simplest realization of such a model is considered here, in order to bring out at the qualitative level the microstructure effect on the magnetization curve. 
This model is treated in the framework of micromagnetism, {\it i.e.} through a continuous medium type of
approach, and the calculations are performed by using the \mbox{MAGPAR} \cite{magpar} code based on the finite elements method.
%
      
\section{Experiment}
A polycrystalline PrCo$_3$ was prepared by melting high-purity starting elements ($>99.9\%$) in an induction furnace under a protective environment(Ar atmosphere). Their homogeneity was insured with five times consecutive melting. These ingots were used as pre-alloys to manufacture samples by high-energy milling. All powder handling was performed in a glove box under a high- purity Ar atmosphere with an O$_2$ and H$_2$O rate around 1\,ppm with vial sealed glove box. 

The mixtures were sealed in hardened steel vial together with 15\,mm diameter steel balls. Milling was performed for 5\,h under a high purity Ar atmosphere with a ball to powder ratio of $15:1$, in a high-energy planetary ball mill Fritsch P7. These milling conditions correspond to kinetic shock energy, shock frequency and injected shock power values respectively equal to 0.81\,J/hit, 62\,Hz and 19.5\,watt/g. After milling, the powder mixtures was wrapped in tantalum foil and sealed in silica tubes under a vacuum of 
10$^{-6}$\,Torr, 
then annealed for 30\,min at temperature between 750 and 1050$^\circ$C followed by quenching in water.

The purity of the final phase was checked by X-ray diffraction (XRD) using Brucker diffractometer mounted with the Bragg-Brentano geometry and automatic divergence slit, using the CuK$\alpha $ radiation. The unit parameters were measured with Si as standard ($a$= 5.4308\,\AA) to insure a unit-cell parameter within the accuracy of $\pm$1$\times $10$^{-3}$\,\AA. The counting rate was 22\,s per scanning step and step size of 0.04$^\circ$.
To determine the microstructural alloy properties (crystallite size) and crystallographic properties (phase identification and lattice parameters), XRD patterns have been analyzed by Rietveld method.

The data treatment was carried out by a Rietveld refinement \cite{r67,r69} as implemented in the \mbox{FULLPROF} computer code \cite{rfm91, r93} with the assumption of a peak line profile of Thompson-Cox-Hastings allowing multiple phase refinement of each of the coexisting phases \cite{rc91}. It takes into account the broadening of the diffraction lines and gives, via the Scherrer formula, the autocoherent diffraction domain size. The weight percentage of each of the coexisting phases is directly obtained  with the approximation of a Brindley absorption factor equal to 1. The goodness-of-fit indicators R$_B$ and $\chi ^2$ are calculated as usual and described earlier \cite{db00, bd01}.    
 
The magnetic anisotropy of PrCo$_3$ was studied using X-ray diffraction. The X-ray diffraction pattern was recorded on powder sample (grain size $< 50\,\mu$m) oriented under an external magnetic field. The magnetic field was oriented perpendicular to the surface of the sample and the grains are oriented and shape the needles parallel to the field applied, the powder so directed is analyzed by XRD at room temperature, thus the x-ray pattern gave information about the easy magnetization direction.

Curie temperature T$_c$ was 
measured on a differential sample magnetometer \mbox{MANICS} in field of $0.1$\,kOe and heating rate of 10\,K/min with around $10$\,mg sample sealed under vacuum in silica tube in order to prevent oxidation under heating. The Curie temperature was determined from the $M-T$ curves by extrapolating linear part of the $M-T$ curves of the intersection with the extended base line. 
The $M-H$ curves (Magnetic hysteresis measurement) were plotted at T= 10 and 293\,K with a Physical Properties Measurement System (PPMS) Quantum Design equipment and a maximum applied field of 90\,kOe on sample in epoxy resin.   

\section{Results and discussion}

\subsection{Structure analysis}
\begin{figure}[h!]
	\centering
    \includegraphics*[width= 4cm]{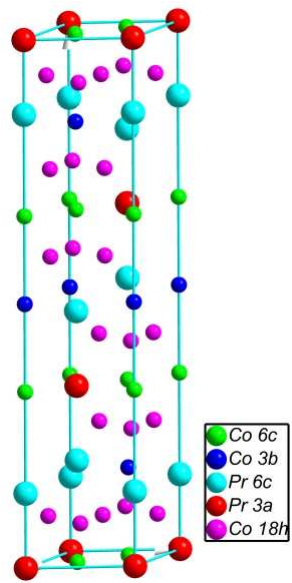} 
    \caption{\label{fig:1} Crystal structure of PrCo$_3$. Pr atoms occupy the 3$a$ and 6$c$ sites, and Co atoms occupy the 3$b$, 6$c$ and 18$h$ sites.}
\end{figure}

The PrCo$_3$ compounds crystallize in rhombohedral PuNi$_3$-type structure (space group: $R\bar3m$). The crystal lattice of PrCo$_3$, can be regarded, approximately, as consisting of the consecutive arrangement of alternating blocks of PrCo$_5$ and PrCo$_2$ along the $c$ axis with the ration $1:2$ (Fig.~\ref{fig:1}), having almost the same structure as in the corresponding compounds PrCo$_5$ and PrCo$_2$. The unit cell contains two non-equivalent crystallographic site for Pr ions, $3a$ and $6c$, and three sites for Co: $3b$, $6c$ and $18h$. 

\begin{figure}[h!]
	\centering
    {\centering \resizebox*{3.3in}{!}{\includegraphics*{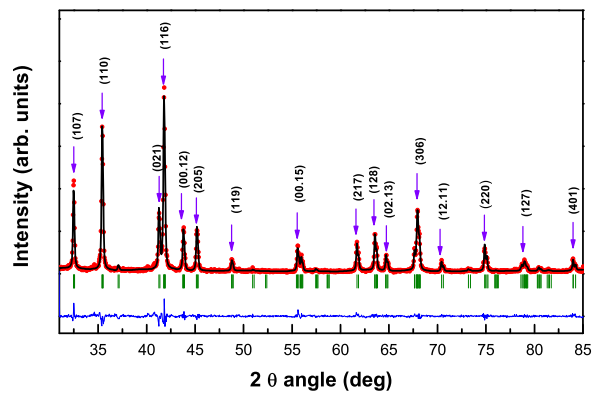}} \par}
    \caption{\label{fig:2} Rietveld analysis for PrCo$_3$ ingot.}
\end{figure}

The X-ray diffraction (XRD) of the PrCo$_3$ shows that the compound is single phase with the rhombohedral PuNi$_3$-type structure (space group: $R\bar3m$). Fig.~\ref{fig:2} presents the analysis result of XRD pattern of PrCo$_3$ ingot using the Rietveld analysis. The lattice parameters of PrCo$_3$ compound are
$a$= 5.069\,\AA\ and $c$= 24.795\,\AA. The atomic positions, R$_B$, $\chi^2$ factors from Rietveld fit are given in Table 1. These values of structural parameters are in agreement with the results obtained previously by \cite{bky79}. This ingot is an appropriate precursor for using as pre-alloys to manufacture samples by high-energy milling. 

\begin{figure}[h!]
	\centering
    {\centering \resizebox*{3.2in}{!}{\includegraphics*{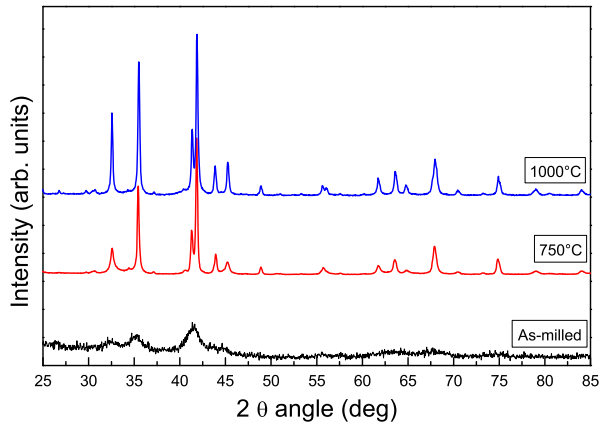}} \par} 
 \caption{\label{fig:3} XRD pattern of the PrCo$_3$ as-milled and annealed samples at 750 and 1000$^\circ$C for 30\,min.}
\end{figure}

\begin{table}[bth!]
\begin{center}
\caption{\label{tab:table1}
$a$ and $c$ cell parameters, atomic positions, R$_B$ and $\chi^2$ factors from Rietveld fit.} 
\begin{tabular}{lc}
\hline \hline
a(\AA)       &  5.0690   \\
c(\AA)       & 24.7950   \\
$z(6c)$Pr    &  0.1414   \\
$z(6c)$Co    &  0.3337   \\
$x(18h)$Co   &  0.5002   \\
$y(18h)$Co   &  0.4998   \\
$z(18h)$Co   &  0.0829   \\
$c/a$        &  4.8915   \\   
V(\AA)$^3$   &  551.42   \\
R$_B$        &  5.050   \\
$\chi^2$     &  2.300   \\
\hline
\hline
\end{tabular}
\end{center}
\end{table}
Figure~\ref{fig:3} 
shows the x-ray diffraction pattern of three samples of PrCo$_3$, the as-milled powder and annealed at 750 and 1000$^\circ$C for 30\,min. 

The as-milled powder is typical of a quasiamorphous phase and does not show characteristic lines of diffraction of the crystallized state, however, we observe in the low angles a flattened and large line around 42$^\circ$. Such quasiamorphous matrix suggests the presence of various types of small crystallites.
The diffraction peaks become more pronounced with increasing annealing temperature. The annealed samples at 750 and 1000$^\circ$C for 30\,min reveal the presence of PrCo$_3$ phases crystallized. We have observed a small amount (around 0.5\,wt$\%$) of Pr$_2$O$_3$ in all samples annealed after milling.

The mean diffraction crystallite size of the powder after annealing at 750 and 1000$^\circ$C for 30\,min obtained by fitting the XRD diagram with Rietveld method, ranges between 31 and 45\,nm respectively, this implies that synthesized materials are noncrystalline.   

\subsection{Intrinsic magnetic properties}
The Curie temperature, also called the magnetic ordering temperature is a direct measure of the exchange interaction, which is the origin of ferromagnetism. This interaction depends markedly on the interatomic distance. 

Generally, the Curie temperature in rare-earth transition-metal intermetallic compounds is governed by three kinds of exchange interactions, namely, the $3d-3d$ exchange interaction $(J_{\text{CoCo}})$ between the magnetic moments of the Co sublattice, $4f-4f$ exchange interaction $(J_{\text{PrPr}})$ between the magnetic moments within the Pr sublattice, and the intersublattice $3d-4f$ exchange interaction $(J_{\text{PrCo}})$. The interactions between the rare-earth spins $4f-4f$ are assumed to be weak and negligible in comparison with the other two types of interactions \cite{gl74}.
There are two exchange interactions in PrCo$_3$ compounds, positive and negative. When the distance of the Co-Co pairs is smaller than 2.45\,\AA, the exchange interactions is negative, whereas at larger Co-Co distances the interaction is positive.

\begin{figure}[h!]
	\centering
    {\centering \resizebox*{3.1in}{!}{\includegraphics*{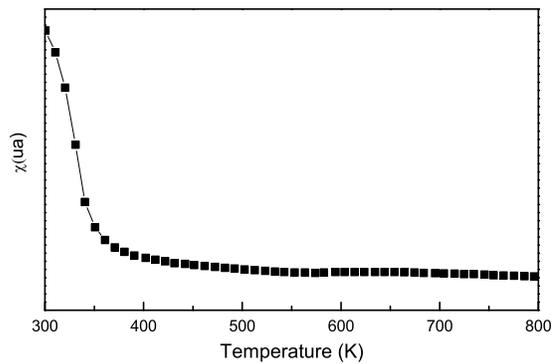}} \par}
    \caption{\label{fig:4} Thermomagnetization curves of PrCo$_3$ obtained after high-energy milling and annealed at 750$^\circ$C for 30\,min.}
\end{figure}

As an example, the thermomagnetic magnetization curve of PrCo$_3$ high-energy milling sample annealed at 750$^\circ$C for 30\,min is presented in Fig.~\ref{fig:4}. This sample has a unique magnetic phase transition at the Curie temperature of 349\,K. This result is in agreement with the data given by Lemaire \cite{l66}.

The low Curie temperature in PrCo$_3$ compounds is due to the short Co-Co interatomic distances at the sites $6c-18h$ and at $18h-18h$ (Table 2), where the Co atoms couple antiferromagnetically. In this structure, the $6c-18h$ and at $18h-18h$ interactions are strongly negative.

\begin{table}
\begin{ruledtabular}
\caption{\label{tab:table2}
Interatomic distance (\AA) for PrCo$_3$.}
\begin{tabular}{cccccc}
      Site & Co $(3b)$ & Co $(6c)$ & Co $(18h)$ &  Pr $(6a)$  &  Pr $(6c)$ \\
          \hline 
          \hline          
 	Co $(3b)$	&   5.07 & 4.12 & 2.54 & 2.99 & 5.07  \\
	Co $(6c)$	&   4.12 & 2.92 & 2.42 & 4.56 & 2.92  \\
	Co $(18h)$ 	&   2.53 & 2.42 & 2.41 & 2.92 & 3.50  \\
	Pr $(6a)$ 	&   2.99 & 4.56 & 2.99 & 3.18 & 3.50  \\
    Pr $(6c)$ 	&   2.99 & 2.92 & 3.26 & 3.50 & 5.07  \\
\end{tabular}
\end{ruledtabular}
\end{table}  

\begin{figure}[h!]
	\centering
    {\centering \resizebox*{3.1in}{!}{\includegraphics*{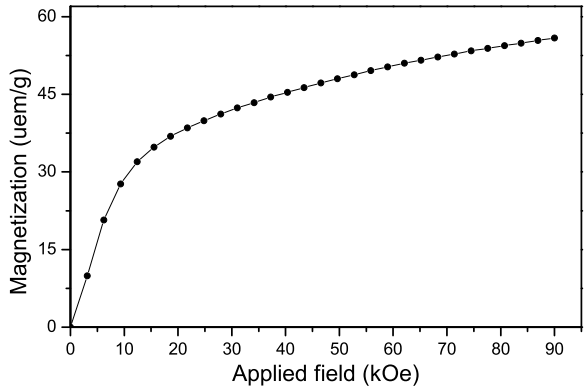}} \par}
    \caption{\label{fig:5} Magnetization curves of PrCo$_3$ obtained after high-energy milling and annealed at 750$^\circ$C for 30\,min, measured at 293\,K.}
\end{figure}

Isothermal magnetization curves $M-H$ obtained at 293\,K is represented in Fig.~\ref{fig:5} for PrCo$_3$. This figure shows that at 293\,K the saturation is not reached, which gives evidence for a magnetocrystalline anisotropy among the highest known in those kinds of systems.

\begin{figure}[h!]
	\centering
    {\centering \resizebox*{3.1in}{!}{\includegraphics*{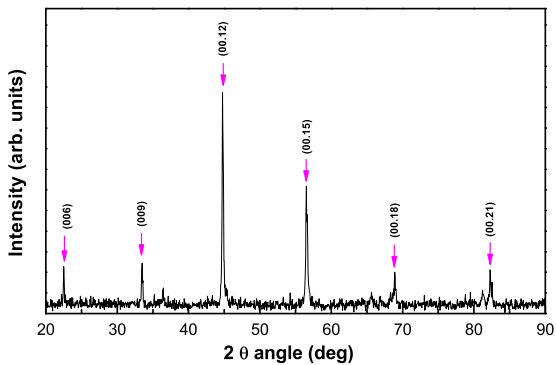}} \par}
     \caption{\label{fig:6}  XRD pattern of PrCo$_3$ oriented.}
\end{figure}

The pattern obtained on sample oriented under an external magnetic field, applied perpendicularly to the plane of the sample, shows only $(00l)$ Bragg peaks (Fig.~\ref{fig:6}), which amount to suggest that the anisotropy strengthens the peaks of diffraction of type $(00l)$ and at room temperature the easy magnetization direction is parallel to the c-axis. A similar result is obtained for YCo$_{3-x}$Fe$_x$ by \cite{bgk00} obtained by oriented powder under extern field.   
\subsection{Extrinsic magnetic properties}
\begin{figure}[h!]
	\centering
    {\centering \resizebox*{3.0in}{!}{\includegraphics*{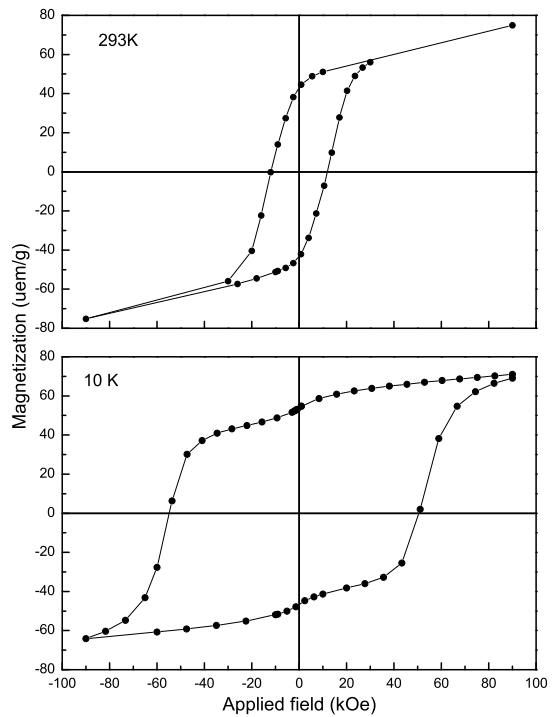}} \par}
    \caption{\label{fig:7} Hysteresis loop of PrCo$_3$ milling before and after annealing at 750$^\circ$C for 30\,min, measured at 293 and 10\,K.}
\end{figure}

Concerning the extrinsic magnetic properties, 
we have measured hysteresis loops at 10 and 293\,K for all samples annealed at different temperature (750 to 1050$^\circ$C for 30\,min) and plotted $M-H$ curves. 
Fig.~\ref{fig:7} show as an 
example the hysteresis loops measured at 10 and 293\,K for sample annealed at 750$^\circ$C. This sample exhibits high coercivities, we have obtained, for $T= 10$\,K, $H_\text{c}= 55$\,kOe, $M_\text{R}= 56$\,uem/g and $M_\text{R}/M_\text{max}= 0.75$. For T= 293\,K, $H_\text{c}= 12$\,kOe and $M_\text{R}= 44$\,uem/g and $M_\text{R}/M_\text{max}= 0.60$ inherent to nanocrystalline state. The coercivity of 55\,kOe at 10\,K confirms 
the large magnetocrystalline anisotropy. 

\section{Simulation} 

In order to get a first picture of the link between the structure at the nanometric
scale and the extrinsic magnetic properties at least at the qualitative level,
we have performed micromagnetic simulations on a model for our system.
Indeed, micromagnetic simulations provide a useful tool for this purpose
and have been done for a wide range of nanocomposite magnetic systems
\cite{nano_comp_1, nano_comp_2, nano_comp_3, grif_01, rev_skomski, nano_layer_1}.
It is worth mentioning that quite generally such simulations deal with
models for heterogeneous systems including magnetic grains of different materials,
typically hard and soft materials, the purpose being to characterize the exchange 
\cite{nano_comp_1, nano_comp_2, nano_comp_3, grif_01} 
or the magnetostatic coupling between phases \cite{dip_int_1}. 
The general finding is that the inter-grains exchange coupling leads to a
decrease of the coercivity and an increase of remanence when compared to the 
values of the hard material component. The magnetostatic coupling may be
more complex due to the long ranged and the anisotropy nature of dipolar interactions
\cite{dip_int_2}.
Here, we model a biphasic system,
the phases of which are of the same chemical composition.
Indeed, as we have seen in the experimental section, the as milled sample is
amorphous, and then after annealing well crystallized grains whose size is
controlled by the annealing temperature are formed. 
Our purpose, rather than to provide a systematic investigation, 
is to draw out some of the salient features concerning the 
sample structure which may explain the shape of the magnetization curve.
We model the experimental sample as an assembly of well crystallized grains,
represented by either spherical or cubic inclusions,
embedded in the matrix made of the amorphous phase corresponding to the as 
milled material.
The inclusion shape should be discriminated on physical grounds related
to the crystallization growth at the temperature considered. This is beyond the
scope of the present work, and we will be lead to discriminate the grain shape 
from the agreement with experimental results. We emphasize that the main difference 
between these two cases is the larger relative importance taken by the amorphous matrix,
playing the role of grain boundaries, in the model with spherical inclusions.
 
Since the crystallites are well crystallized, they are characterized
by the magnetic constants of the perfect massive crystal.
We take the experimental value of the saturation magnetization per unit volume, 
\mbox{$J_s= 0.723$\,T} as deduced
from both the saturation magnetization per unit formula \cite{l66}, namely
3.8$\mu_B$ and the volumic mass as obtained from the measured volume of the unit cell
(see table \ref{tab:table1}) which corresponds to $Z= 9$.
The value of the exchange constant is set equal to that of the archetype of hard material 
(Nd$_2$Fe$_{14}$B), namely \mbox{$A_{ex}= 7.7\times 10^{-12}$\,J/m} and the magnetocrystalline anisotropy constant
is fitted on the experimental value of the coercive field we have got at \mbox{T= 10\,K},
with the result \mbox{$K_1= 3.05~10^6$\,J/m$^3$}.
The amorphous matrix is on the one hand non anisotropic ($K_1= 0$), and on the other hand
characterized by a very small value of the exchange constant, set somewhat arbitrarily to
$A_{ex}= 3.8\times 10^{-13}\,$J/m.
Indeed, since the coercivity remains very high we expect in our two phase model only a 
negligible exchange coupling between the matrix and the inclusions. 
Then, concerning the matrix saturation magnetization, we expect a much smaller value 
than for the crystallized phase of the inclusions \cite{sab}; therefore we consider either a 
non magnetic (\mbox{$J_s$= 0}) or only weakly polarized 
(\mbox{$0.05\;T \leq J_s \leq 0.25\;T$}) matrix.
As we shall see these two last points can be justified
by the magnetization curve shape.
The inclusions have all the same size, while their easy axis, $\hat{n}_i$
are distributed either randomly or in a solid angle \mbox{$\Theta \leq \Theta_M$}
centered on the direction of the external field, $\hat{h}_{ext}$ in order to model
the effect of the texturation by the field. 
The influence of the preferential
orientation has been investigated only in the case of spherical inclusions.
The inclusions are located on the sites of simple cubic lattice.
The inclusion packing fraction is therefore limited to values smaller than
$\pi/6$ in the case of spherical inclusions; we set the nearest neighbor distance $d= 2.106\;R$
close to the minimum value $d= 2~R$, which leads to $\varphi= 0.445$.
We have not such limitation when dealing with cubic inclusions. In this later case
the inclusions are separated by a grain boundary of thickness, say $\delta$, and
the volume fraction is then \mbox{$\varphi= (1 + \delta/2R)^{-3}$} where the inclusion hedge
length is set to $2~R$. In the framework of the micromagnetism, the value of $\delta$ cannot be 
reduced at will since  $\delta$ must not reach the 
microscopisc scale; here we consider either \mbox{$\delta/d= 0.10$} or \mbox{$\delta/d= 0.237$}
where $d$ is the minimum intergrain distance which leads to $\varphi~=~0.73$ or 0.445
respectively. 
The embedding matrix is a parallepipedic box including 4 layers of 64 lattice sites and the
total number of particles taken into account is thus 256.
The applied field is set normal to the layers.
In order to simulate as far 
as possible the experimental case, we have chosen \mbox{$R= 17\,$nm} for the spherical 
inclusions and \mbox{$2\;R= 32.5$\,nm} for the cubic inclusions. Accordingly
the size of the matrix is \mbox{$288\,$nm $\times 288\,$nm $\times 144\,$nm}
and in between \mbox{$297\,$nm $\times 297\,$nm $\times 158\,$nm}
and \mbox{$340\,$nm $\times 340\,$nm $\times 170\,$nm} in the case of spherical and cubic 
inclusions respectively.
The micromagnetic calculations are performed by using the \mbox{MAGPAR}
code \cite{magpar}, based on a
finite elements scheme, from either the LLG equation or the total energy minimization.
Most of the calculations 
have been performed with a mesh including 1632000 tetrahedra,
and we have checked on a refined mesh of 3159000 tetrahedra the convergence versus the
element size of the results in the case of spherical inclusions.

\begin{figure}[h!]
{\centering \resizebox*{3.2in}{!}{\includegraphics*{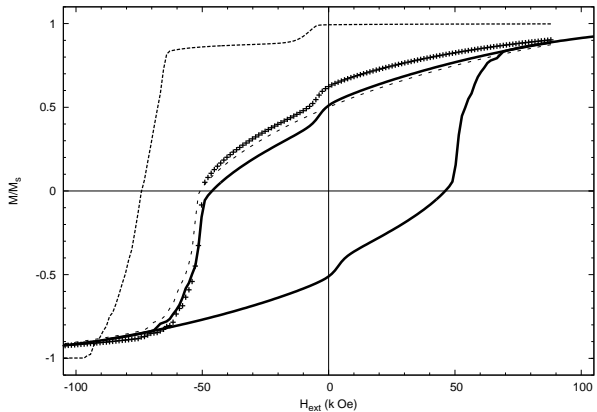}} \par}
\caption{ \label{fig_simu_sph}
Calculated de-magnetization curve in terms of the external field
for the systems of spherical inclusions with $\varphi~=0.45$.
Randomly distributed axes and matrix saturation magnetization
$J_s= 0$, dashed line; $J_s= 0.05$\,T, solid line.
The whole hysteresis curve in this later case is obtained by using the symmetric
of the de-magnetization curve.
$J_s = 0.05$\,T and $\Theta_M= \pi/2.3$, crosses; $\Theta_M= \pi/15$, dotted line.
}
\end{figure}

The results for the model with spherical inclusions are displayed in figure (\ref{fig_simu_sph}).
First of all, we consider the case of the non magnetic matrix ($J_s= 0$) with randomly
oriented particles. As expected, the coercivity takes a large value, namely totally governed
by that of the particles, since the exchange constant of the matrix is too low for the
particles to be exchanged coupled to and {\it via} the matrix. We have checked that even with 
$J_s= 0$ a significant value of the exchange constant for the amorphous matrix leads to
a decrease of the coercivity, and this as already mentioned justifies our choice of
a very small value of $A_{ex}$ for the matrix. However, the main qualitative characteristic
which is missing in this first calculated magnetization curve when compared to the experimental 
one is the small kink in the vicinity of zero external field. Such a kink may be the signature 
of an uncoupled two phase magnetic behavior. Hence, we have increased the value of the matrix
saturation magnetization in order to get such a behavior and we find that \mbox{$J_s= 0.05$\,T}
leads to a qualitative agreement with the experimental curve. 
When the matrix $J_s$ value increases the kink amplitude becomes much larger than the 
experimental one. As a matter of fact, the kink relative amplitude behaves roughly as 
\begin{eqnarray}
\frac{(1 - \varphi) J_s^{(mat)}}{(1 - \varphi) J_s^{(mat)} + \varphi J_s^{(incl)}}
\nonumber
\end{eqnarray}
%

\begin{figure}[h!]
{\centering \resizebox*{3.2in}{!}{\includegraphics*{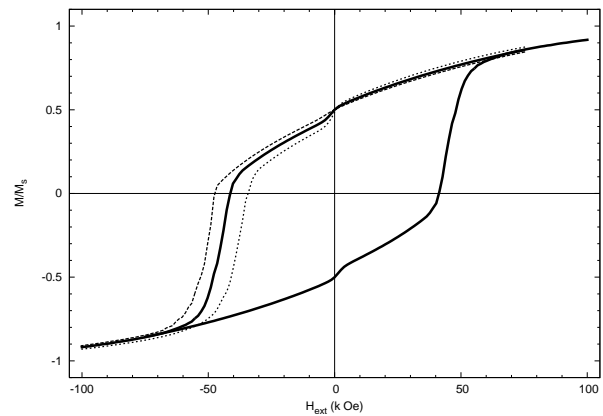}} \par}
\caption{ \label{fig_simu_cub}
Calculated de-magnetization curve in terms of the external field
for the systems of cubic inclusions with $\varphi= 0.73$.
Randomly distributed axes and matrix saturation magnetization
$J_s= 0.05$\,T, dashed line; $J_s= 0.125$\,T, solid line and $J_s= 0.25$\,T, dotted line.
}
\end{figure}

The agreement between the calculated hysteresis curve and the experimental one, 
for randomly distributed axes and $J_s^{(mat)}= 0.05\,$T is qualitatively 
satisfying. We have to mention however, that we display in figure (\ref{fig_simu_sph}) 
the magnetization curve reduced by the total saturation magnetization, $M_s$,
of the system the value of which, 
\mbox{$M_s= \varphi J_s^{(incl)} + (1 - \varphi) J_s^{(mat)}$} 
is significantly smaller than the experimental one. This is due both to the 
small value of the saturation magnetization of the amorphous matrix
obtained to reproduce the amplitude of the kink at $H_{ex}= 0$, and 
to the large fraction of the total volume occupied by the matrix. 
The model with cubic inclusions can remedy to this drawback since then the
inclusion packing fraction is not limited as already mentioned. The results obtained
with \mbox{$\varphi= 0.73$} are displayed on figure (\ref{fig_simu_cub}) for the matrix 
saturation 
magnetization $J_s^{mat}$= 0.05\,T, 0.125\,T and 0.250\,T. The first conclusion is that at the
qualitative level, the hysteresis curve looks like that of the system with spherical
inclusions which may be the consequence of the easy axis random orientation.
Now, since the matrix volume fraction is strongly reduced, the amplitude
of the magnetisation kink due to the matrix magnetization reversal compares to the experimental
one with a larger value for $J_s^{mat}$ than in the model with spherical inclusions, namely 
\mbox{$J_s^{mat} \sim$ 0.125\,T}. However, increasing the matrix saturation magnetization
leads to an important matrix/inclusion magnetostatic coupling resulting in a non negligible
reduction of the coercivity (see figure \ref{fig_simu_cub}). Thus the best agreement with
the experimental results is obtained with the model with cubic inclusions, $\varphi= 0.73$
and $J_s^{(mat)}= 0.125$\,T ({\it i.e.} $J_s^{(mat)}/J_s^{(incl)}= 0.172$).
With this set of parameters, we get for the saturation magnetization of the composite
system \mbox{$J_s^{(tot)}= 0.777\;J_s^{(incl)}$}.
The simulated magnetization curve of the model with cubic inclusions and $\varphi= 0.445$ is
quite close to the that of the model with spherical inclusions and is thus not shown.
We also get an important reduction of the coercivity with the increase of the matrix
saturation magnetization.

Then, and only for the model with spherical inclusions, we have examined the effect of the texturation 
{\it via} the easy axes distribution. In this calculation, the maximum value of the polar angle $\Theta$
of the axes {$\hat{n}_i$} is either $\pi/15$ or $\pi/2.3$ for a strong or a very weak
preferential orientation respectively. Notice 
the strong sensitivity to the variation of $\Theta_M$; indeed the random distribution corresponds to 
$\Theta_M= \pi/2$. We emphasize that this orientation effect may represent a consequence of a non 
spherical shape for the particles. Finally, we have also calculated the magnetization curve for particles
of size \mbox{$R= 35\,$nm} with nearly the same result at least in the case on a non magnetic matrix. 
Therefore, as expected, we conclude that the strong size effect observed on the coercivity of 
nanocomposite materials is mainly due to the dependence of the crystallite magnetic properties on the size. 

\section{Conclusion}
In summary, the PrCo$_3$ intermetallic compound, crystallize in rhombohedral PuNi$_3$-type structure (space group: $R\bar3m$). This compound is a highly anisotropic uniaxial ferromagnet with the easy magnetization direction is parallel to the c-axis. No saturation reached at room temperature for applied field of 90\,kOe. High coercivities of about 12\,kOe at 293\,K and 55\,kOe at 10\,K can be reached in PrCo$_3$ alloy prepared by high-energy milling method followed by annealing at 750$^\circ$C for 30\,min. High energy milled method following the controlled nanocrystallization brings high magnetic properties with smaller grain size more adequate to favor an available microstructure to perform this properties, so these results show that PrCo$_3$compounds are promising materials for permanent magnet applications. 
From the model calculations of the macroscopic extrinsic properties we can propose a simple picture of the microstructure. Indeed, the magnetization curve presents a signature of a system made of two uncoupled phases: the magnetically hard crystallites and the amorphous matrix.
The presence of the small kink in the magnetization curve is analyzed as the reversal of
the matrix magnetization; the localization of this kink at nearly zero applied field and the
its small amplitude are indications that the matrix is, as expected, magnetically very soft and
characterized by a small saturation magnetization. This latter point must be taken with care, since
here we propose a simple lattice for the crystallite which may lead to an overestimated fraction of 
the total volume occupied by the matrix also with the model with cubic inclusions. It is important
to notice that when dealing with the reduced magnetization, $M/M_s$ we find that the influence 
of the grain shape is rather weak.

\section*{Acknowledgements}
Most of the calculations have been performed at the CINES (Centre Informatique National de l'Enseignement
Sup\'erieur).


\begin{thebibliography}{25}
\expandafter\ifx\csname natexlab\endcsname\relax\def\natexlab#1{#1}\fi
\expandafter\ifx\csname bibnamefont\endcsname\relax
  \def\bibnamefont#1{#1}\fi
\expandafter\ifx\csname bibfnamefont\endcsname\relax
  \def\bibfnamefont#1{#1}\fi
\expandafter\ifx\csname citenamefont\endcsname\relax
  \def\citenamefont#1{#1}\fi
\expandafter\ifx\csname url\endcsname\relax
  \def\url#1{\texttt{#1}}\fi
\expandafter\ifx\csname urlprefix\endcsname\relax\def\urlprefix{URL }\fi
\providecommand{\bibinfo}[2]{#2}
\providecommand{\eprint}[2][]{\url{#2}}

\bibitem[{\citenamefont{Gutfleisch et~al.}(2004)\citenamefont{Gutfleisch,
  Dempsey, Yan, M{\"u}ller, and D.Givord}}]{gdy04}
\bibinfo{author}{\bibfnamefont{O.}~\bibnamefont{Gutfleisch}},
  \bibinfo{author}{\bibfnamefont{N.~M.} \bibnamefont{Dempsey}},
  \bibinfo{author}{\bibfnamefont{A.}~\bibnamefont{Yan}},
  \bibinfo{author}{\bibfnamefont{K.~H.} \bibnamefont{M{\"u}ller}},
  \bibnamefont{and} \bibinfo{author}{\bibnamefont{D.Givord}},
  \bibinfo{journal}{J. Magn. Magn. Mater.} \textbf{\bibinfo{volume}{272}},
  \bibinfo{pages}{647} (\bibinfo{year}{2004}).

\bibitem[{\citenamefont{McCormick et~al.}(1996)\citenamefont{McCormick, Ding,
  Feutrill, and Street}}]{mdf96}
\bibinfo{author}{\bibfnamefont{P.~G.} \bibnamefont{McCormick}},
  \bibinfo{author}{\bibfnamefont{J.}~\bibnamefont{Ding}},
  \bibinfo{author}{\bibfnamefont{E.~H.} \bibnamefont{Feutrill}},
  \bibnamefont{and} \bibinfo{author}{\bibfnamefont{R.}~\bibnamefont{Street}},
  \bibinfo{journal}{J. Magn. Magn. Mater.} \textbf{\bibinfo{volume}{157}},
  \bibinfo{pages}{7} (\bibinfo{year}{1996}).

\bibitem[{\citenamefont{Rivera-G{\'o}mez
  et~al.}(2009)\citenamefont{Rivera-G{\'o}mez, Elizalde-Galindo, and
  Matutes-Aquino}}]{rem09}
\bibinfo{author}{\bibfnamefont{F.~J.} \bibnamefont{Rivera-G{\'o}mez}},
  \bibinfo{author}{\bibfnamefont{J.~T.} \bibnamefont{Elizalde-Galindo}},
  \bibnamefont{and} \bibinfo{author}{\bibfnamefont{J.~A.}
  \bibnamefont{Matutes-Aquino}}, \bibinfo{journal}{J. Alloys Compd.}
  \textbf{\bibinfo{volume}{477}}, \bibinfo{pages}{588} (\bibinfo{year}{2009}).

\bibitem[{\citenamefont{Chen et~al.}(2000)\citenamefont{Chen, Meng-Burany,
  Okumura, and Hadjipanayis}}]{cmhh00}
\bibinfo{author}{\bibfnamefont{Z.}~\bibnamefont{Chen}},
  \bibinfo{author}{\bibfnamefont{X.}~\bibnamefont{Meng-Burany}},
  \bibinfo{author}{\bibfnamefont{H.}~\bibnamefont{Okumura}}, \bibnamefont{and}
  \bibinfo{author}{\bibfnamefont{G.~C.} \bibnamefont{Hadjipanayis}},
  \bibinfo{journal}{J. Appl. Phys.} \textbf{\bibinfo{volume}{87}},
  \bibinfo{pages}{3409} (\bibinfo{year}{2000}).

\bibitem[{\citenamefont{Scholz et~al.}(2003)\citenamefont{Scholz, Schrefl,
  Suess, Dittrich, Forster, and Tsiantos}}]{magpar}
\bibinfo{author}{\bibfnamefont{W.}~\bibnamefont{Scholz}},
  \bibinfo{author}{\bibfnamefont{T.}~\bibnamefont{Schrefl}},
  \bibinfo{author}{\bibfnamefont{D.}~\bibnamefont{Suess}},
  \bibinfo{author}{\bibfnamefont{R.}~\bibnamefont{Dittrich}},
  \bibinfo{author}{\bibfnamefont{H.}~\bibnamefont{Forster}}, \bibnamefont{and}
  \bibinfo{author}{\bibfnamefont{V.}~\bibnamefont{Tsiantos}},
  \bibinfo{journal}{Comput. Mater. Sci.} \textbf{\bibinfo{volume}{28}},
  \bibinfo{pages}{366} (\bibinfo{year}{2003}).

\bibitem[{\citenamefont{Rietveld}(1967)}]{r67}
\bibinfo{author}{\bibfnamefont{H.}~\bibnamefont{Rietveld}},
  \bibinfo{journal}{Acta Crystallogr} \textbf{\bibinfo{volume}{22}},
  \bibinfo{pages}{151} (\bibinfo{year}{1967}).

\bibitem[{\citenamefont{Rietveld}(1969)}]{r69}
\bibinfo{author}{\bibfnamefont{H.}~\bibnamefont{Rietveld}},
  \bibinfo{journal}{J. Appl. Crystallogr} \textbf{\bibinfo{volume}{2}},
  \bibinfo{pages}{65} (\bibinfo{year}{1969}).

\bibitem[{\citenamefont{Rodriguez-Carvajal
  et~al.}(1991{\natexlab{a}})\citenamefont{Rodriguez-Carvajal, Fernandez-Diaz,
  and Martinez}}]{rfm91}
\bibinfo{author}{\bibfnamefont{J.}~\bibnamefont{Rodriguez-Carvajal}},
  \bibinfo{author}{\bibfnamefont{M.~T.} \bibnamefont{Fernandez-Diaz}},
  \bibnamefont{and} \bibinfo{author}{\bibfnamefont{J.~L.}
  \bibnamefont{Martinez}}, \bibinfo{journal}{J. Phys.: Condens. Matter}
  \textbf{\bibinfo{volume}{3}}, \bibinfo{pages}{3215}
  (\bibinfo{year}{1991}{\natexlab{a}}).

\bibitem[{\citenamefont{Rodriguez-Carvajal}(1993)}]{r93}
\bibinfo{author}{\bibfnamefont{J.}~\bibnamefont{Rodriguez-Carvajal}},
  \bibinfo{journal}{Physica B} \textbf{\bibinfo{volume}{192}},
  \bibinfo{pages}{55} (\bibinfo{year}{1993}).

\bibitem[{\citenamefont{Rodriguez-Carvajal
  et~al.}(1991{\natexlab{b}})\citenamefont{Rodriguez-Carvajal, Fernandez-Diaz,
  and Martinez}}]{rc91}
\bibinfo{author}{\bibfnamefont{J.}~\bibnamefont{Rodriguez-Carvajal}},
  \bibinfo{author}{\bibfnamefont{M.~T.} \bibnamefont{Fernandez-Diaz}},
  \bibnamefont{and} \bibinfo{author}{\bibfnamefont{J.~L.}
  \bibnamefont{Martinez}}, \bibinfo{journal}{J. Phys.: Condens. Matter}
  \textbf{\bibinfo{volume}{3}}, \bibinfo{pages}{3215}
  (\bibinfo{year}{1991}{\natexlab{b}}).

\bibitem[{\citenamefont{Djega-Mariadassou and Bessais}(2000)}]{db00}
\bibinfo{author}{\bibfnamefont{C.}~\bibnamefont{Djega-Mariadassou}}
  \bibnamefont{and} \bibinfo{author}{\bibfnamefont{L.}~\bibnamefont{Bessais}},
  \bibinfo{journal}{J. Magn. Magn. Mater.} \textbf{\bibinfo{volume}{210}},
  \bibinfo{pages}{81} (\bibinfo{year}{2000}).

\bibitem[{\citenamefont{Bessais and Djega-Mariadassou}(2001)}]{bd01}
\bibinfo{author}{\bibfnamefont{L.}~\bibnamefont{Bessais}} \bibnamefont{and}
  \bibinfo{author}{\bibfnamefont{C.}~\bibnamefont{Djega-Mariadassou}},
  \bibinfo{journal}{Phys. Rev. B} \textbf{\bibinfo{volume}{63}},
  \bibinfo{pages}{054412} (\bibinfo{year}{2001}).

\bibitem[{\citenamefont{Burnasheva et~al.}(1979)\citenamefont{Burnasheva,
  Klimeshin, Yartys', and Semenenko}}]{bky79}
\bibinfo{author}{\bibfnamefont{V.~V.} \bibnamefont{Burnasheva}},
  \bibinfo{author}{\bibfnamefont{V.~V.} \bibnamefont{Klimeshin}},
  \bibinfo{author}{\bibfnamefont{V.~A.} \bibnamefont{Yartys'}},
  \bibnamefont{and} \bibinfo{author}{\bibfnamefont{K.~N.}
  \bibnamefont{Semenenko}}, \bibinfo{journal}{Inorganic Materials, translated
  from Izvestiya Akademii Nauk SSSR, Neorganicheskie Materialy}
  \textbf{\bibinfo{volume}{15}}, \bibinfo{pages}{627} (\bibinfo{year}{1979}).

\bibitem[{\citenamefont{Givord and Lemaire}(1974)}]{gl74}
\bibinfo{author}{\bibfnamefont{D.}~\bibnamefont{Givord}} \bibnamefont{and}
  \bibinfo{author}{\bibfnamefont{R.}~\bibnamefont{Lemaire}},
  \bibinfo{journal}{IEEE Trans. Magn.} \textbf{\bibinfo{volume}{MAG-10}},
  \bibinfo{pages}{109} (\bibinfo{year}{1974}).

\bibitem[{\citenamefont{Lemaire}(1966)}]{l66}
\bibinfo{author}{\bibfnamefont{R.}~\bibnamefont{Lemaire}},
  \bibinfo{journal}{Cobalt} \textbf{\bibinfo{volume}{33}}, \bibinfo{pages}{201}
  (\bibinfo{year}{1966}).

\bibitem[{\citenamefont{Bartashevich et~al.}(2000)\citenamefont{Bartashevich,
  Goto, and Koui}}]{bgk00}
\bibinfo{author}{\bibfnamefont{M.~I.} \bibnamefont{Bartashevich}},
  \bibinfo{author}{\bibfnamefont{T.}~\bibnamefont{Goto}}, \bibnamefont{and}
  \bibinfo{author}{\bibfnamefont{K.}~\bibnamefont{Koui}},
  \bibinfo{journal}{Physica B} \textbf{\bibinfo{volume}{292}},
  \bibinfo{pages}{9} (\bibinfo{year}{2000}).

\bibitem[{\citenamefont{Fischer et~al.}(1995)\citenamefont{Fischer, Schrefl,
  Kronm\"uller, and Fidler}}]{nano_comp_1}
\bibinfo{author}{\bibfnamefont{R.}~\bibnamefont{Fischer}},
  \bibinfo{author}{\bibfnamefont{T.}~\bibnamefont{Schrefl}},
  \bibinfo{author}{\bibfnamefont{H.}~\bibnamefont{Kronm\"uller}},
  \bibnamefont{and} \bibinfo{author}{\bibfnamefont{J.}~\bibnamefont{Fidler}},
  \bibinfo{journal}{J. Magn. Magn. Mater.} \textbf{\bibinfo{volume}{150}},
  \bibinfo{pages}{329} (\bibinfo{year}{1995}).

\bibitem[{\citenamefont{Fischer and Kronm\"uller}(1998)}]{nano_comp_2}
\bibinfo{author}{\bibfnamefont{R.}~\bibnamefont{Fischer}} \bibnamefont{and}
  \bibinfo{author}{\bibfnamefont{H.}~\bibnamefont{Kronm\"uller}},
  \bibinfo{journal}{J. Magn. Magn. Mater.} \textbf{\bibinfo{volume}{184}},
  \bibinfo{pages}{166} (\bibinfo{year}{1998}).

\bibitem[{\citenamefont{Zheng and Zhao}(2009)}]{nano_comp_3}
\bibinfo{author}{\bibfnamefont{B.}~\bibnamefont{Zheng}} \bibnamefont{and}
  \bibinfo{author}{\bibfnamefont{S.}~\bibnamefont{Zhao}}, \bibinfo{journal}{J.
  Rare Earths} \textbf{\bibinfo{volume}{27}}, \bibinfo{pages}{145}
  (\bibinfo{year}{2009}).

\bibitem[{\citenamefont{Griffiths et~al.}(2001)\citenamefont{Griffiths, Bichop,
  Tucker, and Davies}}]{grif_01}
\bibinfo{author}{\bibfnamefont{M.}~\bibnamefont{Griffiths}},
  \bibinfo{author}{\bibfnamefont{J.}~\bibnamefont{Bichop}},
  \bibinfo{author}{\bibfnamefont{J.}~\bibnamefont{Tucker}}, \bibnamefont{and}
  \bibinfo{author}{\bibfnamefont{H.}~\bibnamefont{Davies}},
  \bibinfo{journal}{J. Magn. Magn. Mater.} \textbf{\bibinfo{volume}{234}},
  \bibinfo{pages}{331} (\bibinfo{year}{2001}).

\bibitem[{\citenamefont{Skomski}(2003)}]{rev_skomski}
\bibinfo{author}{\bibfnamefont{R.}~\bibnamefont{Skomski}}, \bibinfo{journal}{J.
  Phys.: Condens. Matter} \textbf{\bibinfo{volume}{15}}, \bibinfo{pages}{R841}
  (\bibinfo{year}{2003}).

\bibitem[{\citenamefont{Rong et~al.}(2006{\natexlab{a}})\citenamefont{Rong,
  Zhang, Chen, Shen, and He}}]{nano_layer_1}
\bibinfo{author}{\bibfnamefont{C.-B.} \bibnamefont{Rong}},
  \bibinfo{author}{\bibfnamefont{H.-W.} \bibnamefont{Zhang}},
  \bibinfo{author}{\bibfnamefont{R.-J.} \bibnamefont{Chen}},
  \bibinfo{author}{\bibfnamefont{B.-G.} \bibnamefont{Shen}}, \bibnamefont{and}
  \bibinfo{author}{\bibfnamefont{S.-L.} \bibnamefont{He}}, \bibinfo{journal}{J.
  Appl. Phys.} \textbf{\bibinfo{volume}{100}}, \bibinfo{pages}{123913}
  (\bibinfo{year}{2006}{\natexlab{a}}).

\bibitem[{\citenamefont{Rong et~al.}(2006{\natexlab{b}})\citenamefont{Rong,
  Zhang, Chen, He, and Shen}}]{dip_int_1}
\bibinfo{author}{\bibfnamefont{C.-B.} \bibnamefont{Rong}},
  \bibinfo{author}{\bibfnamefont{H.-W.} \bibnamefont{Zhang}},
  \bibinfo{author}{\bibfnamefont{R.-J.} \bibnamefont{Chen}},
  \bibinfo{author}{\bibfnamefont{S.-L.} \bibnamefont{He}}, \bibnamefont{and}
  \bibinfo{author}{\bibfnamefont{B.-G.} \bibnamefont{Shen}},
  \bibinfo{journal}{J. Magn. Magn. Mater.} \textbf{\bibinfo{volume}{302}},
  \bibinfo{pages}{126} (\bibinfo{year}{2006}{\natexlab{b}}).

\bibitem[{\citenamefont{Russier}(2009)}]{dip_int_2}
\bibinfo{author}{\bibfnamefont{V.}~\bibnamefont{Russier}}, \bibinfo{journal}{J.
  Appl. Phys.} \textbf{\bibinfo{volume}{105}}, \bibinfo{pages}{073915}
  (\bibinfo{year}{2009}).

\bibitem[{\citenamefont{Sab et~al.}(2003)\citenamefont{Sab, Bessais,
  Dj\'ega-Mariadassou, Dan, and Phuc}}]{sab}
\bibinfo{author}{\bibfnamefont{S.}~\bibnamefont{Sab}},
  \bibinfo{author}{\bibfnamefont{L.}~\bibnamefont{Bessais}},
  \bibinfo{author}{\bibfnamefont{C.}~\bibnamefont{Dj\'ega-Mariadassou}},
  \bibinfo{author}{\bibfnamefont{N.}~\bibnamefont{Dan}}, \bibnamefont{and}
  \bibinfo{author}{\bibfnamefont{N.}~\bibnamefont{Phuc}}, \bibinfo{journal}{J.
  Phys.: Condens. Matter} \textbf{\bibinfo{volume}{15}}, \bibinfo{pages}{5615}
  (\bibinfo{year}{2003}).

\end{thebibliography}

\end{document}